\documentclass[pra,twocolumn,aps,showpacs,preprintnumbers,amsmath,amssymb,superscriptaddress]{revtex4-1}

\usepackage{graphicx}
\usepackage{dcolumn}
\usepackage{bm}
\usepackage{amsmath}
\usepackage{cancel}
\usepackage{color}

\begin{document}


\title{Cooper pair polaritons in cold fermionic atoms within a cavity}

\author{Amaury Dodel}
\affiliation{State Key Laboratory of Precision Spectroscopy, School of Physical and Material Sciences,
East China Normal University, Shanghai 200062, China}
\affiliation{New York University Shanghai, 1555 Century Ave, Pudong, Shanghai 200122, China}

\author{Alexander Pikovski}
\affiliation{New York University Shanghai, 1555 Century Ave, Pudong, Shanghai 200122, China}

\author{Igor Ermakov}
\affiliation{ITMO University, Kronverkskiy 49, 197101, St.Petersburg, Russia}
\affiliation{New York University Shanghai, 1555 Century Ave, Pudong, Shanghai 200122, China}

\author{Marek Narozniak}
\affiliation{New York University Shanghai, 1555 Century Ave, Pudong, Shanghai 200122, China}
\affiliation{Department of Physics, New York University, New York, NY 10003, USA}

\author{Valentin Ivannikov}
\affiliation{New York University Shanghai, 1555 Century Ave, Pudong, Shanghai 200122, China}

\author{Haibin Wu}
\affiliation{State Key Laboratory of Precision Spectroscopy, School of Physical and Material Sciences,
East China Normal University, Shanghai 200062, China}
\affiliation{NYU-ECNU Institute of Physics at NYU Shanghai, 3663 Zhongshan Road North, Shanghai 200062, China}

\author{Tim Byrnes}
\affiliation{New York University Shanghai, 1555 Century Ave, Pudong, Shanghai 200122, China}
\affiliation{State Key Laboratory of Precision Spectroscopy, School of Physical and Material Sciences, East China Normal University, Shanghai 200062, China}
\affiliation{NYU-ECNU Institute of Physics at NYU Shanghai, 3663 Zhongshan Road North, Shanghai 200062, China}
\affiliation{National Institute of Informatics, 2-1-2 Hitotsubashi, Chiyoda-ku, Tokyo 101-8430, Japan}
\affiliation{Department of Physics, New York University, New York, NY 10003, USA}

\date{\today}

\begin{abstract}
We formulate a Bardeen-Cooper-Schriffer (BCS) theory of quasiparticles in a degenerate Fermi gas strongly coupled to photons in a optical cavity. The elementary photonic excitations of the system are cavity polaritons, which consist of a cavity photon and an excitation of an atom within the Fermi sea.  The excitation of the atom out of the Fermi sea leaves behind a hole, which together results in a loosely bound Cooper pair, allowing for the system to be written by a BCS wavefunction.  As the density of the excitations is increased, the excited atom and hole become more strongly bound, crossing over into the molecular regime.  This thus realizes an alternative BCS to BEC crossover scenario, where the participating species are quasiparticle excitations in a Fermi sea consisting of excited atoms and holes.  
\end{abstract}

\maketitle

\section{Introduction} 

Degenerate atomic gases are a thriving platform for the realization of highly controllable quantum many-body systems.  Using combinations of various techniques developed over the last few decades, it is possible to create a wide range of quantum systems.  For example, using optical lattices and taking advantage of the natural interactions one can realize the Bose-Hubbard model and observe a quantum phase transition between a Mott insulator and a superfluid \cite{greiner2002m}.  For fermionic atoms, the crossover between a molecular Bose-Einstein condensate (BEC) and Bardeen-Cooper-Schriffer (BCS) phase was observed by tuning the Feshbach resonance \cite{regal2004observation}.  Since these pioneering works, numerous examples of other physical models have been realized, ranging from those with other lattice geometries, spin-orbit coupling, and artificial gauge fields \cite{buluta2009quantum,georgescu2014quantum,bloch2012quantum,gross2017quantum}.  The general approach, dubbed ``quantum simulation'' allows one to realize and study quantum-many body systems in engineered, rather than naturally occuring systems.  Due to the difficulty of theoretically and numerically studying such systems, this offers a new route towards understanding such system, realizing Feynman's conjecture that quantum mechanical systems are suited towards the simulation of other quantum systems \cite{feynman1982simulating}.  It also offers opportunities to create and study systems that do not occur naturally to observe novel physics.

Spurred on by the observation of BECs in cold atomic gases, it has been of great interest to reproduce this general effect in other physical systems. To date, other systems where BECs have been observed are with magnons \cite{nikuni00,demokritov06}, photons \cite{klaers10}, and exciton-polaritons \cite{kasprzak06,deng02}. In particular, exciton-polaritons are a hybrid quasiparticle in a semiconductor consisting of a superposition of a cavity photon and an exciton \cite{deng10,byrnes14}.  Excitons are bound pairs of electrons and holes, and are created by optical excitations between the valence and conduction bands, and possess a repulsive interaction \cite{ciuti1998role,byrnes2014effective}.   Similarly to atomic BECs, polariton BECs can be manipulated in various ways such as applying periodic potentials of various geometries \cite{lai07,kim11,masumoto12}.  One of the interesting aspects of exciton-polariton BECs is that they can be observed at much higher temperatures than for atoms, ranging from 10K to room temperature.

Another forefront in recent years has been the development of interfaces between quantum gases and photons by putting cold atoms in cavities \cite{ritsch2013cold}.  The first realizations of this were using degenerate bosons put in optical cavities \cite{brennecke2007cavity,colombe2007strong}, where strong coupling between photons and the excitations of the BECs were observed.    In addition to optical cavities, hybrid systems with BECs and superconducting resonators have also been realized \cite{verdu2009strong,bernon2013manipulation}.  For optical cavities working in the dispersive regime \cite{gupta2007cavity}, this can form the foundation of a long-ranged interaction mediated by the photons \cite{larson2008cold,mottl2012roton,maschler2005cold,colella2018quantum,schlawin2019cavity}. More recently, several works have investigated the physics of degenerate Fermi gases in an optical cavity \cite{chen2014superradiance,kanamoto2010optomechanics,larson2008cold,guo2012ultracold}, predicting various effects such as fermionic superradiance \cite{keeling2014fermionic}, chiral states \cite{sheikhan2016cavity}, artificial magnetic fields \cite{kollath2016ultracold}, quantum phase transitions \cite{muller2012quantum,colella2018quantum}, and topological edge states \cite{schlawin2019cavity}.

\begin{figure}
\includegraphics[width=\columnwidth]{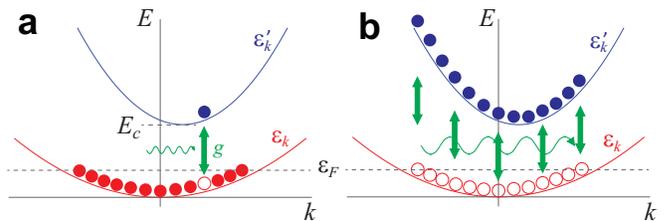}
\caption{Schematic figure of the (a) low and (b) high excitation density regimes of degenerate fermions in a optical cavity. At low density most of the atoms occupy the internal ground state which has an energy-momentum dispersion $ \epsilon_{\bm{k}} = \frac{\hbar^2 k^2}{2m}  $. The polariton consists of a superposition of a cavity photon and an excited state atom with a dispersion $\epsilon_{\bm{k}}' = \frac{\hbar^2 | \bm{k} + \bm{k}_0 |^2}{2m'} $, which leaves behind a hole in the Fermi sea. At high excitation density, all the atoms participate and are in a 50/50 superposition of the ground and excited states. }
\label{fig1}
\end{figure}

In this paper, we consider a degenerate Fermi gas with two internal levels coupled to an optical cavity. The excitations of an optical cavity containing atoms is well-known to be described by vacuum Rabi splitting of the energy levels of the atoms, where cavity polaritons are the elementary excitations 
\cite{agarwal1984vacuum,carmichael1991classical,raizen1989normal,gripp1996anharmonicity,tuchman2006normal,wu2008observation,colombe2007strong,culver2016collective}. Such a system has almost an exact analogy to exciton-polaritons in semiconductor systems.  The degenerate gas of fermions in the ground state plays the role of the valence band, and the excited state plays the role of the conduction band  (Fig. \ref{fig1}).  A photon excites an atom to the excited state, leaving behind a hole.  There are however also several differences of the cavity polaritons to exciton-polaritons, namely for semiconductors supporting excitons the direct band gap means that the dispersion of the valence band has negative effective mass (positive curvature of the valence band).  In addition, the fermions are charged electrons and holes, hence have a Coulomb attraction between them.

One of the main results of this paper is that it is possible to realize a novel type of BEC-BCS crossover, in terms of the 
the excited atoms and holes in the Fermi sea.  This forms a different type of BEC-BCS crossover that are {\it quasiparticles} within the background Fermi sea, rather than the atoms themselves as has been considered before \cite{regal2004observation,guo2012ultracold}.  To see in what sense a BCS state is possible, consider the excitation of an atom in the Fermi sea by a photon.  In this case, if the photon has a momentum $ \bm{k}_0 $, this can potentially excite an excited atom with momentum $ \bm{k}_0 +  \bm{k} $ and produce a hole moving with momentum $ - \bm{k} $.   Thus the relative momentum between the excited atom and hole is not bound by momentum conservation, and can form a wavefunction that is describable by a BCS state.   Due to the effect of the cavity and the interactions, we show that the excited atoms and holes have an attractive force mediated by the photons. In the regime that it is energetically favorable for the excited atom and hole to be loosely bound, a Cooper pair of these fermions form.  The strong coupling of the Cooper pair to light via the cavity creates a ``Cooper pair polariton''.  These Cooper pair polaritons are closely related to what is observed in semiconductor systems \cite{byrnes2010bcs,kamide2010determines,comte1982exciton,imamouglu1998phase}, but the physics of these systems are yet to be observed to our knowledge. One major difference to the analogous case in semiconductors is that the BCS regime is much more easily observed, due to the lack of Coulomb attraction between excited atoms and holes.  This makes the Cooper pair polaritons far more easily observed, whereas for excitons the electrons and holes are generally always strongly bound.

\section{Formalism for fermionic atoms in a cavity} 

\subsection{Theoretical Model}

We consider a gas of two-level fermionic atoms in a cavity, described by the Hamiltonian
\begin{align}
H & = \sum_{\bm{k}} \left[ \frac{\hbar^2 k^2}{2m}   b^\dagger_{\bm{k}} b_{\bm{k}} + (E_c + \frac{\hbar^2 k^2}{2m'}  ) c^\dagger_{\bm{k}} c_{\bm{k}}  + \hbar \omega_{\bm{k}}  a^\dagger_{\bm{k}} a_{\bm{k}} \right]   \nonumber \\
& + \sum_{\bm{k} \bm{k}' } \left[    g  c^\dagger_{\bm{k}+\bm{k}'}  b_{\bm{k}}  a_{\bm{k}'} +  g^* a^\dagger_{\bm{k}'}  b_{\bm{k}}^\dagger c_{\bm{k}+\bm{k}'}  \right] \nonumber \\
& + \sum_{\bm{k} \bm{k}' \bm{q}}  V_{bc} b^\dagger_{\bm{k}+\bm{q}} c^\dagger_{\bm{k}'-\bm{q}} c_{\bm{k}'} b_{\bm{k}} .
\label{ham0}
\end{align}
Here, $ b_{\bm{k}}, c_{\bm{k}} $ are the fermionic annihilation operators for the ground and excited state of the atoms, and $ a_{\bm{k}} $ is the bosonic annihilation operator for the cavity mode with momentum $ \bm{k} $.  For generality, we consider the mass of the atoms for the ground and excited state atoms respectively to be given by $ m, m' $ respectively.  
$ E_c $ is the energy difference between the internal states of the fermions, $ g $ is the cavity-atomic transition coupling, $ V_{bc} $ is the energy densities for the inter-species interaction between the ground and excited states.  We henceforth consider that the cold fermions are placed in a ring cavity with a resonance momentum $ \bm{k}_0 $ with magnitude $ k_0 \equiv  | \bm{k}_0 | = 2\pi/\lambda_0 $, where $ \lambda_0 $ is the wavelength of light.  The ring cavity simplifies our analysis such that the light propagation is in a single direction.  

For optical cavities that are tuned to the atomic resonance, the cavity photon momentum $ k_0 $ and the Fermi momentum $ k_F $ of the atoms are of a similar order.  Their energy dispersions are on the other hand highly mismatched. To see this, consider the energy of a photon in the vicinity of the cavity photon resonance
\begin{align}
\hbar \omega_{\bm{k}} & = \hbar c | \bm{k} | = \hbar c | \bm{k}_0 + \delta \bm{k} | \nonumber \\
& = \hbar k_0 c  \sqrt{ 1 + \frac{ 2 \bm{k}_0 \cdot \delta \bm{k} +  \delta k^2}{k_0^2}} \nonumber \\
& \approx \hbar k_0 c  + \frac{ \hbar c}{k_0} (\bm{k}_0+ \frac{\delta \bm{k}}{2} )\cdot \delta \bm{k} \nonumber \\
& = \frac{\hbar k_0 c}{2} + \frac{ \hbar^2 | \bm{k}_0 + \delta \bm{k} |^2}{2 m_{\text{ph}}}
\end{align}
where we defined $ \delta \bm{k} = \bm{k} - \bm{k}_0 $ and expanded the square root in the second last line. The effective photon mass is $ m_{\text{ph}} = \hbar k_0 / c $, which for typical atomic parameters is much lighter than the atomic mass $ m_{\text{ph}}  \ll m, m' $.   
This means that the only relevant photons that couple with the atoms are those with momentum $ \delta \bm{k} = 0 $ (i.e. $ \bm{k} = \bm{k}_0 $), which are tuned on the atomic resonance, and all other photons will be highly off-resonant.  The extremely light effective photon mass means that the polaritons also have an extremely light effective mass \cite{byrnes14}.  In the same way as exciton-polaritons, this provides the conditions for macroscopic condensation of polaritons.  

Taking this into account, the Hamiltonian then can be 
\begin{align}
H & = \sum_{\bm{k}} \left[ \epsilon_{\bm{k}}  b^\dagger_{\bm{k}} b_{\bm{k}} + (E_c + \epsilon_{\bm{k}}' ) c^\dagger_{\bm{k}} c_{\bm{k}} \right]  + \hbar \omega  a^\dagger a  \nonumber \\
& +\sum_{\bm{k}} \left[    g  c^\dagger_{\bm{k}}  b_{\bm{k}}  a +  g^* a^\dagger b_{\bm{k}}^\dagger c_{\bm{k}}   \right] + \sum_{\bm{k} \bm{k}' \bm{q}}  V_{bc} b^\dagger_{\bm{k}+\bm{q}} c^\dagger_{\bm{k}'-\bm{q}} c_{\bm{k}'} b_{\bm{k}} .
\label{ham}
\end{align}
where we have changed the labeling of the excited atoms according to $ c_{\bm{k}+ \bm{k}_0}  \rightarrow c_{\bm{k}} $ and abbreviated the photon operator as $ a_{\bm{k}_0}  \rightarrow a $.  The dispersions of the ground and excited state atoms are $ \epsilon_{\bm{k}} = \frac{\hbar^2 k^2}{2m} $, $ \epsilon_{\bm{k}}' = \frac{\hbar^2 | \bm{k} + \bm{k}_0 |^2}{2m'} $ respectively, and $ \hbar \omega $ is the cavity photon energy.   

We also define the atom number and excitation number operators
\begin{align}
\hat{N} & = \sum_{\bm{k}} b^\dagger_{\bm{k}} b_{\bm{k}} + \sum_{\bm{k}} c^\dagger_{\bm{k}} c_{\bm{k}} \nonumber \\
 n_{\text{ex}} & = a^\dagger a + \sum_{\bm{k}} c^\dagger_{\bm{k}} c_{\bm{k}},
\end{align}
which commute with the Hamiltonian $ [H, N] = [H, n_{\text{ex}}] = 0 $, and thus has a common set of eigenstates.  We assume that the total number of atoms is $ N = \sum_{|\bm{k}|\le k_F } 1 $ and $  k_F  $ defines the Fermi momentum for the non-interacting system ground state.

We note that the intra-species interaction never contributes for momentum independent interactions in (\ref{ham0}) due to fermionic statistics 
\begin{align}
 & \sum_{\bm{k} \bm{k}' \bm{q}}  b^\dagger_{\bm{k}+\bm{q}} b^\dagger_{\bm{k}'-\bm{q}} b_{\bm{k}'} b_{\bm{k}} \nonumber \\
& =  \frac{1}{2}  \sum_{\bm{k} \bm{k}' \bm{q}}  b^\dagger_{\bm{k}+\bm{q}} b^\dagger_{\bm{k}'-\bm{q}} b_{\bm{k}'} b_{\bm{k}} + 
\frac{1}{2}  \sum_{\bm{k} \bm{k}' \bm{q}}  b^\dagger_{\bm{k}+\bm{q}} b^\dagger_{\bm{k}'-\bm{q}} b_{\bm{k}} b_{\bm{k}'} 
= 0 
\end{align}
where we have made a transformation of indices $ \bm{k} \rightarrow \bm{k}' $,  $ \bm{k}' \rightarrow \bm{k} $, $ \bm{q} \rightarrow - \bm{k}' + \bm{k} + \bm{q} $ in the second term.

\subsection{BCS mean field theory}

We now construct a BCS mean field theory which can describe both the low and high excitation regime of the cavity-Fermi gas system.   We perform a standard BCS mean-field theory such that only quadratic terms in the Hamiltonian are present, with a ground state of the form (see Appendix for details)
\begin{align}
| \Phi \rangle = \prod_{\bm{k}} \left[ u_{\bm{k}} + v_{\bm{k}} c^\dagger_{\bm{k}} h^\dagger_{-\bm{k}} \right] |E_0 (k_F) \rangle  ,
\label{bcswavefunction}
\end{align}
where $ |u_{\bm{k}}|^2 +  |v_{\bm{k}}|^2 = 1 $ and $ |E_0 (k_F) \rangle = \prod_{| \bm{k}|  < k_F}  b^\dagger_{\bm{k}} |0 \rangle $ is the Fermi sea, and we have defined fermionic hole operators as $ h_{\bm{k}} = b_{-\bm{k}}^\dagger 
$ \cite{yamamoto1999mesoscopic}.  In order that this state is the ground state of the mean-field Hamiltonian, the BCS parameters must obey 
\begin{align}
2  \delta \xi_{\bm{k}}  u_{\bm{k}} v_{\bm{k}}- \Delta ( |v_{\bm{k}}|^2 -  |u_{\bm{k}}|^2 ) = 0 .
\label{bcsquadratic}
\end{align}
Solving the quadratic equation we find the BCS equations
\begin{align}
v_{\bm{k}}^2 & = \langle c^\dagger_{\bm{k}} c_{\bm{k}} \rangle = \frac{1}{2} \left(
1- \frac{ \delta \xi_{\bm{k}}}{ E_{\bm{k}}} \right) \label{v2relation} \\
u_{\bm{k}} v_{\bm{k}} & = \langle c^\dagger_{\bm{k}} h^\dagger_{-\bm{k}} \rangle = \frac{\Delta}{2 E_{\bm{k}}}
\end{align}
where 
\begin{align}
 \delta \xi_{\bm{k}} & = (\xi_{\bm{k}}'  - \xi_{\bm{k}})/2 , \nonumber \\
 \xi_{\bm{k}}  & = \epsilon_{\bm{k}} + V_{bc} X , \nonumber \\
\xi_{\bm{k}}'  & = E_c + \epsilon_{\bm{k}}'  + (N +1- X) V_{bc} - \mu \nonumber \\  
\Delta & = V_{bc} D - g \lambda^* \nonumber \\ 
 E_{\bm{k}} & = \sqrt{ (\delta \xi_{\bm{k}})^2 + \Delta^2} .
\end{align}
The mean field values are 
\begin{align}
 X & = \sum_{\bm{k}} \langle h^\dagger_{\bm{k}} h_{\bm{k}} \rangle = \sum_{\bm{k}} \langle c^\dagger_{\bm{k}} c_{\bm{k}} \rangle, \nonumber \\
 D & = \sum_{\bm{k}} \langle c^\dagger_{\bm{k}} h^\dagger_{-\bm{k}} \rangle, \nonumber \\  
 \lambda & = \langle a \rangle .
\end{align}
These are solved self-consistently to find the parameters $ v_{\bm{k}}, u_{\bm{k}} $.

\section{Solutions in limiting cases}

\subsection{Low excitation limit} 

The BCS formalism allows us to smoothly interpolate between the low excitation and high excitation regimes.   Let us first obtain some limiting cases of the theory. In the low excitation regime we expect $ \lambda, |v_{\bm{k}}| \ll 1 $, and thus we can approximate (\ref{bcsquadratic}) as \cite{comte1982exciton}
\begin{align}
2  \delta \xi_{\bm{k}}  v_{\bm{k}}- ( V_{bc} +\frac{|g|^2}{\hbar \omega - \mu}  ) \sum_{\bm{k}'} v_{\bm{k}'}
= 0 ,
\label{lowdensitylimit}
\end{align}
where  $\mu $ is a  chemical potential to control the number of excitations in the system. The above equation is a Schrodinger equation in momentum space with a delta-function attractor with respect to the wavefunction $ v_{\bm{k}} $  \cite{byrnes2010bcs}. 
The cavity provides provides an effective attractive force between the excited atoms and holes.  The strength of the interaction $ |g|^2/(\hbar \omega - \mu) $ suggests that it is a second order energy shift, in a similar way to the ac Stark shift or Feshbach resonance.  Given the presence of an excited atom and hole, the cavity can induce a relaxation and re-excitation back to the original state, giving rise to an energy shift.  In addition to the cavity, the inter-species repulsion $ V_{bc} $ can also  produce an effective attractive interaction.  The physical intuition for this is that for a repulsive inter-species interaction, it is energetically favorable for the excited atom to recombine with the hole, annihilating each other.  In order for this to happen, the excited atom and hole must have matching momenta, according to the cavity coupling term $ b^\dagger_{\bm{k}} c_{\bm{k}} = h_{-\bm{k}}c_{\bm{k}} $ in the Hamiltonian (\ref{ham}), thereby producing an effective attractive force between them.

In three dimensions the bound states take the form
\begin{align}
 v_{\bm{k}} \propto \frac{1}{k^2 + \alpha^2}
\label{lowdenswavefunc}
\end{align}
where $ \alpha $  is the parameter that is determined by the boundary conditions.  In three dimensions there are an infinite number of bound states, making the evaluation of $ \alpha $ problematic using standard techniques \cite{geltman2011bound}.  For the self-consistent theory, we can however fix $ \alpha $ by demanding that the product $ \delta \xi_{\bm{k}}  v_{\bm{k}} $ in (\ref{lowdensitylimit}) is independent of $ k $.  For example, when $ k_0 = 0 $ we have the condition $ \frac{\hbar^2 \alpha^2}{2 \tilde{m}} = E_c + V_{bc}(N+1) - \mu $, where  $ 1/\tilde{m} = 1/m'-1/m $.  The normalization of $  v_{\bm{k}} $ is fixed by the number of excited atoms, according to (\ref{v2relation}). 

We thus write the polariton creation operator
\begin{align}
p^\dagger  = \chi  a^\dagger + \frac{\zeta}{\sqrt{N_0}} 
\sum_{\bm{k}} v_{\bm{k}}  c^\dagger_{\bm{k}}  h_{-\bm{k}}^\dagger,
\label{polaritonoperator}
\end{align} 
where $\chi, \zeta $ are coefficients that satisfy $ |\chi|^2 + |\zeta|^2 = 1 $, and 
$ N_0 = \sum_{\bm{k}} | v_{\bm{k}} |^2 $. This operator obeys approximate bosonic commutation relations, as can be verified by evaluating the commutator
\begin{align}
[p, p^\dagger  ] = 1 - \frac{|\zeta|^2}{N_0} \sum_{\bm{k}} | v_{\bm{k}} |^2 ( c^\dagger_{\bm{k}} c_{\bm{k}} + h^\dagger_{-\bm{k}} h_{-\bm{k}}  ).
\end{align}
As long as the number of excited atomic states (and hence holes) is small, the second term above gives a small contribution, and follows bosonic commutation relations.  For the case that the ground and excited state dispersions are the same $ \epsilon_{\bm{k}}' =\epsilon_{\bm{k}} $, it is possible to explicitly write the exact polariton wavefunction for low densities without a mean-field approximation, corresponding to $  v_{\bm{k}} = 1 $.  In this case applying the operator (\ref{polaritonoperator}) on the Fermi sea gives an exact eigenstate of (\ref{ham}) with excitation energy (see Appendix)
\begin{align}
E_p = \frac{E_c + \hbar \omega}{2} \mp \sqrt{\frac{(E_c - \hbar \omega)^2}{4} + |g|^2 N } .
\label{polaritonenergy}
\end{align}
The Rabi splitting of the polaritons scales with the square root of the number of atoms as expected.  

The polariton operator (\ref{polaritonoperator}) takes the same form as that familiar with exciton-polaritons in semiconductors \cite{deng10,byrnes14}.  The primary difference here is that in semiconductor systems the electrons and holes are oppositely charged and have a Coulomb attraction, which is lacking in this case. This means that 
for the case of of equal dispersion $ \epsilon_{\bm{k}}' = \epsilon_{\bm{k}} $, there will be no momentum dependence
$ v_{\bm{k}} \propto \text{const} $.  One can view changing the relative dispersion between the ground and excited states as tuning the effective mass in the Schrodinger equation (\ref{lowdensitylimit}) --- a larger relative difference in the dispersions produces a smaller effective mass.  The smaller effective mass produces more strongly bound excited atoms and holes, reducing the momentum dependence in (\ref{polaritonoperator}).  

The BCS parameter $ v_{\bm{k}} $ affects the spontaneous emission rate of the excitations.  Excitons in semiconductors have a much shorter lifetime than free conduction electrons due to superradiant effects \cite{yamamoto1999mesoscopic}.  The excited atom in Fig. \ref{fig1}(a) can only relax to its lower energy state if spatially overlaps with the hole. Using Weisskopf-Wigner theory for the exciton part $ \chi = 0, \zeta = 1 $ of the polariton operator (\ref{polaritonoperator}), we find that the spontaneous emission rate is modified according to 
\begin{align}
\Gamma = \Gamma_0 \left| \frac{1}{\sqrt{N_0}} \sum_{\bm{k}} v_{\bm{k}} \right|^2
\end{align}
where $ \Gamma_0 $ is the spontaneous emission of a single atom.  Depending on the distribution $ v_{\bm{k}}$, the spontaneous emission rate will thus be modified.  For a constant distribution $ v_{\bm{k}}$, corresponding to strongly bound excited atom and holes, the factor $ | \sum_{\bm{k}} v_{\bm{k}}/\sqrt{N_0}|^2  = N $, and we recover Dicke superradiant spontaneous emission.  The opposite limit of $ v_{\bm{k}}  \propto \delta(\bm{k}) $, corresponding to very different dispersions between the ground and excited states gives a factor $  | \sum_{\bm{k}} v_{\bm{k}}/\sqrt{N_0}|^2  = 1 $.  This reduces the spontaneous emission rate due the delocalization of the electron and hole in real space, suppressing the spontaneous emission.

\subsection{High excitation limit}

We now examine the reverse regime when the photon population is large.  In this limit we may set the photonic operator to a classical field $ a \rightarrow  \lambda $ and neglect the interaction terms.  A straightforward diagonalization reveals that the ground state consists of a Fermi sea of a superposition of ground and excited state atoms $ d_{\bm{k}}^\dagger = u_{\bm{k}} b_{\bm{k}}^\dagger + v_{\bm{k}} c_{\bm{k}}^\dagger $.  Such a state can be written in terms of the BCS wavefunction  (\ref{bcswavefunction}).  In the limit of $ \lambda \rightarrow \infty $ the coefficients tend towards $ u_{\bm{k}} = v_{\bm{k}} = 1/\sqrt{2} $.  The energy of an atom  $ d_{\bm{k}}^\dagger  $ in a superposition is
\begin{align}
E_{\bm{k}}^\lambda = & \frac{E_c+ \hbar \omega+ \epsilon_{\bm{k}}' + \epsilon_{\bm{k}}}{2} \nonumber \\
& \pm \sqrt{ \frac{(E_c - \hbar \omega + \epsilon_{\bm{k}}' -\epsilon_{\bm{k}})^2 }{4} +  |g  \lambda|^2} . 
\label{brightcavitylimit}
\end{align}
This takes a similar form to the polariton excitation energy (\ref{polaritonenergy}) except that the  the enhancement factor of the cavity-atom coupling $ g $ scales with the square root of the photon number.  
In addition, in the bright cavity limit the excitations possess momentum dependence. Despite this similarity, we note that the ground state in each limit is rather different, with the bright cavity limit consisting of a superposition of many excited states, whereas the low excitation limit is primarily the undisturbed Fermi sea.

\begin{figure}
\includegraphics[width=\columnwidth]{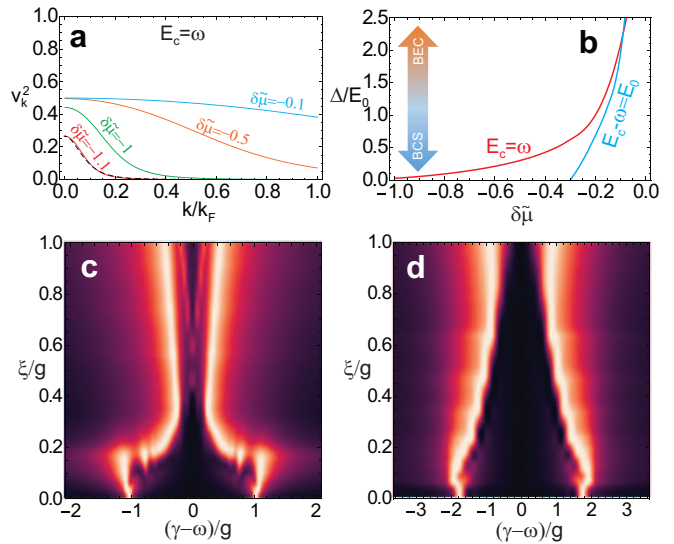}
\caption{Crossover from low to high density for cold fermions in a cavity.  (a)(b)  Solutions of BCS equations found by self-consistent iteration.  (a) The BCS wavefunction parameter $ v_{\bm{k}}^2 $ for zero detuning $ E_c = \omega $. Dashed line shows the low-density solution of the BCS equations (\ref{lowdenswavefunc}) with $ \alpha/k_F \approx 0.12 $.  (b) Same as (a) but for detuning $ E_c - \omega = E_0 $. The BCS gap $ \Delta $ for the detunings as marked. Here we measure all momenta relative to the Fermi momentum $ k_F $ and the is energy scale $ E_0 = \hbar^2 k_F^2/2m $.  Parameters used are $ m = 2m' $, $k_0 = 0 $, $ g/E_0 = 0.01 $, $ V_{bc}/E_0 = 10^{-3} $. The dimensionless chemical potential is defined as $ \delta \tilde{\mu} = 10^3(\mu-\omega)/E_0 $.  (c)(d) The photon amplitude $ | \langle a \rangle |^2 $ in the Tavis-Cummings model with loss and pumping, corresponding to the limit $ V_{bc} = 0 $ and $ \epsilon_{\bm{k}}' = \epsilon_{\bm{k}} $ in the Hamiltonian  (\ref{ham}).  Here $ \xi $ is the pumping strength and $ \gamma $ is the pump frequency.  We take the cavity to be resonant with the atoms $ E_c = \hbar \omega $ and $ \kappa/g = 0.1 $.   The number of atoms and the photon number cutoff is (c) $N=1 $, $ n_{\max} = 20 $; (d) $ N = 3 $, $  n_{\max} = 15 $.  At each pumping $ \xi $ the spectrum is normalized to the same maximum intensity for clarity. } 
\label{fig2}
\end{figure}

\section{Numerical evaluation at intermediate excitations}

\subsection{BCS equations}

In Fig. \ref{fig2} we show the self-consistent evaluation of the BCS equations for various intermediate excitation densities.  We choose parameters consistent with those provided in Ref. \cite{wu2008observation}.  We control the excitation density by varying the difference between the chemical potential $ \delta \mu = \mu- \hbar \omega  $, which must be always a negative quantity since at $ \mu = \hbar \omega $ the photon population becomes infinite. For cavities that are on resonance with the atomic transition $ \hbar \omega = E_c $, in the low-excitation regime we find that the $  v_{\bm{k}} $ distribution follows the predictions of (\ref{lowdenswavefunc}) as shown in Fig. \ref{fig2}(a).  As the excitation density is increased, the distribution flattens out, approaching the high-density limit $  v_{\bm{k}}^2 = u_{\bm{k}}^2 = 1/2 $.  
From (\ref{lowdensitylimit}) it is apparent that the strength of the effective delta-function attractor is larger with a larger photon population, which results in a more strongly bound state between the excited atom and the hole.  In momentum space this results in a broader momentum space distribution for higher excitation densities $ \delta \tilde{\mu} \rightarrow 0^- $.

Another way to see the effect of the attractive force between the excited atom and hole is from the BCS gap energy.  We see in Fig. \ref{fig2}(b) that the gap energy increases as the chemical potential approaches the photon energy. We observe that one can tune the gap energy from a negligible amount, which consist of loosely bound excited atoms and holes, to a large gap energy, consisting of strongly bound excited atoms and holes. The former can be considered to be a BCS-like regime and the latter a BEC regime. The excitation density is therefore one way of controlling the effective strength of attraction between the excited atoms and holes, where the cavity photons act to mediate an attractive force.  Another way is to change the inter-species interaction strength  $V_{bc} $.  From (\ref{lowdensitylimit}) it is apparent that a larger (repulsive) $V_{bc} $ consists of a strong attractive force between excited atoms and holes.  It is interesting to observe that in (\ref{lowdensitylimit}) there is the possibility of the cavity induced energy shift $  |g|^2/(\hbar \omega - \mu)  $ and interaction $V_{bc} $ to cancel each other if the signs are opposed. Thus using a combination of tuning the excitation density, the strength of the inter-species interaction, one can explore a wide range of the BEC-BCS crossover with excited atoms and holes.

\subsection{Cavity spectrum}

Another observable that is experimentally relevant is the spectrum of the cavity in response to photon pumping.  A calculation of the steady-state photon amplitude $ | \langle a \rangle |^2 $ in a Tavis-Cummings model with photon pumping $ \xi $ at frequency $ \gamma $ and photon loss of rate $ \kappa $ is shown in Fig. \ref{fig2}(c)(d).  In using a Tavis-Cummings model we have neglected the atom-atom interactions and assumed that the difference between the ground and excited dispersions play a minimal role (see Appendix). For a single atom limit $ N = 1 $ we see that at low pumping the spectrum corresponds to vacuum Rabi splitting at frequencies $ \gamma = \omega \pm g $. As the pumping increases, the peaks move towards the cavity resonance $ \omega $ and broaden.  For larger number of atoms, we see that at low density the vacuum Rabi splitting at frequencies $ \gamma = \omega \pm g\sqrt{N} $.  Again the peaks move towards the cavity resonance and broaden. This effect of the peaks moving towards the cavity frequency was discussed in several works previously \cite{gripp1996anharmonicity,wu2008observation,wu2009splitting,byrnes2010bcs,ishida2013photoluminescence,ishida2014photoluminescence}. One might naively expect from the result (\ref{brightcavitylimit}) that the Rabi splitting of the system grows from low to high density as  $|g\sqrt{N} | \rightarrow  |g \lambda | $.  This is not the effect that is seen in the cavity spectrum because in the high excitation regime, the state is dominated by a large photonic population, whereas the atomic population is always limited to $ N $.  Hence the state of the system is more dominated by the cavity photons, rather than the atomic states, and the superposition of atom states do not contribute relatively to the 
spectrum \cite{byrnes2010bcs}.

\section{Experimental implementation}

To observe the physics discussed in this paper, the main experimental ingredients in addition to a cold fermion gas at quantum degeneracy are:  (i) a cavity coupling in the strong coupling regime; and (ii) a different dispersion of the ground and excited state atoms  $ \epsilon_{\bm{k}}, \epsilon_{\bm{k}} ' $. The different dispersions can be achieved by having different effective masses of the ground and excited atoms  $ m, m' $, a momentum displacement via the cavity photons $ \bm{k}_0 $, or both.  In order to have differing masses of the ground and excited state atoms, an optical lattice can be used on resonance with either the ground or excited state.  This will modify the mass of the atoms within the optical lattice leaving the other atom unaffected.  We note that the optical lattice should only be made strong enough to modify the dispersion, and the atoms should still be free to move within the lattice, without entering a Mott insulator phase.  With a mass difference between the ground and excited state atoms, the difference between the dispersions takes a typical quadratic form $ \epsilon_{\bm{k}}'- \epsilon_{\bm{k}} \propto | \bm{k} -\bm{k}_0 |^2 $.  
Without the optical lattice, due to the momentum displacement of the cavity photons $ \bm{k}_0 $ there will be a linear momentum dependence of $ \epsilon_{\bm{k}}' - \epsilon_{\bm{k}} = \hbar^2 k_0 ( 2 k + k_0 )/2m $ along the direction of the cavity, which can also be used as the basis of the dispersion in (\ref{lowdensitylimit}).  The cavity transmission can be probed to observe the transition from vacuum Rabi splitting to the cavity resonance energy.   Time-of-flight measurements of the fermion cloud should directly reveal the momentum distribution, where the distribution flattens out at higher densities.   This is similar to the way that the BEC-BCS crossover was detected by tuning the Feshbach resonance between pairings of a cold Fermi gas \cite{regal2005momentum}.  Here the difference is that the cavity has the reverse effect with respect to increasing the density.

The BEC-BCS crossover described in this paper is strongly related to that described in exciton-polariton systems in semiconductors \cite{byrnes2010bcs,kamide2010determines}.  Several experiments in the high-excitation regime for exciton-polaritons have been reported, for example in Refs. \cite{horikiri2016high,horikiri2017highly}. Although in principle these experiments access a similar regime, it has proved difficult so far to conclusively identify these results with the polariton BEC-BCS crossover.  One of the difficulties with semiconductor exciton-polariton systems is that the primary measurement method is via photoluminescence, which occurs due to electron-hole recombination.  This involves the relaxation of the electron in the conduction band to the valence band, and the photon carries the {\it center-of-mass} momentum of the electron and the hole.  Thus the individual (relative) momenta of the electrons and holes are never observed, making it difficult to directly obtain the distribution such as in Fig. \ref{fig2}(a).  The ability to observe the momentum distribution of the atoms directly is a great advantage in atomic systems, in addition to the nearly defect-free realization that is attainable in modern ultracold atomic systems.

\section{Summary and conclusions}

We have shown that the quasiparticle excitations of a cold fermion system coupled to a cavity can be described by a BCS formalism. Depending on the excitation density and atomic interaction, the excited atoms and holes and in the Fermi sea can form either loosely bound Cooper pairs or exciton-like bound states that strongly couple with cavity photons.  
The crossover occurs in the reverse way to the exciton BEC-BCS crossover \cite{comte1982exciton}, where a larger excitation density leads to a stronger attractive force between the excited atoms and holes. The BCS wavefunction approach captures the crossover from the low excitation regime where vacuum Rabi splitting is observed, to the high excitation regime where all atoms are in the superposition of ground and excited states.  Our approach allows one to take into account of the effect of interactions at the same time as the cavity coupling at all excitation densities, hence is a powerful way to track the states of the system across a wide range of parameters.   The BEC-BCS crossover is exactly analogous to the physics of high density exciton-polaritons  \cite{byrnes2010bcs,kamide2010determines} and can be considered an analogue quantum simulator for this system. Cooper pair polaritons are yet to be directly observed due to the difficulty of measurement and other technical complications in semiconductor systems.  Using cold fermions may allow for a simpler route to observing these particles, thanks to measurement techniques which can directly observe the momentum of the atoms, and clean implementation of the system.

\section*{Acknowledgements}

This work is supported by the Shanghai Research Challenge Fund; New York University Global Seed Grants for Collaborative Research; National Natural Science Foundation of China (61571301,D1210036A); the NSFC Research Fund for International Young Scientists (11650110425,11850410426); NYU-ECNU Institute of Physics at NYU Shanghai; the Science and Technology Commission of Shanghai Municipality (17ZR1443600,19XD1423000); the China Science and Technology Exchange Center (NGA-16-001); and the NSFC-RFBR Collaborative grant (81811530112).

\appendix

\section{BCS Theory}

Following the BCS procedure, we start by simplifying the interaction terms in the Hamiltonian such that a Hartree-Fock approximation is made on the intra-species interaction and the inter-species interaction produces transitions on the same momenta (i.e. set $ \bm{q}= \bm{k}'-\bm{k}, 0 $). The Hamiltonian under this approximation then reads
\begin{align}
H &  =  \sum_{\bm{k}} \left[ \epsilon_{\bm{k}}  b^\dagger_{\bm{k}} b_{\bm{k}} + (E_c + \epsilon_{\bm{k}}' + V_{bc} ) c^\dagger_{\bm{k}} c_{\bm{k}} \right]    \nonumber \\
& + \hbar \omega  a^\dagger a +\sum_{\bm{k}} \left[  g  c^\dagger_{\bm{k}}  b_{\bm{k}}  a +  g^* b_{\bm{k}}^\dagger c_{\bm{k}} a^\dagger  \right]   \nonumber \\
& + V_{bc} \sum_{\bm{k} \bm{k}'} \Big[  b^\dagger_{\bm{k}} b_{\bm{k}} c^\dagger_{\bm{k}'} c_{\bm{k}'} 
-   c^\dagger_{\bm{k}} b_{\bm{k}} b^\dagger_{\bm{k}'} c_{\bm{k}'} \Big] ,
\label{hartreeham}
\end{align}
In the above Hamiltonian, we assume that the summation only involves momenta up to the Fermi level $ |\bm{k} | \le k_F $. We justify this simplification by observing that in the low and high excitation limits no atoms are transferred beyond the Fermi momentum.  Hence we expect that at intermediate excitation densities the Fermi surface is also preserved.  

At this point we make a particle-hole transformation, defining fermionic operators $ h_{\bm{k}} = b_{-\bm{k}}^\dagger $ \cite{yamamoto1999mesoscopic}. We thus represent the excitations of the ground state of the atoms in terms of holes in a Fermi sea.  In terms of hole operators, the Hamiltonian now reads
\begin{align}
H &   = 
\sum_{\bm{k}} \Big[ \epsilon_{\bm{k}} - \epsilon_{\bm{k}}  h^\dagger_{\bm{k}} h_{\bm{k}} + (E_c + \epsilon_{\bm{k}}' + (N+1) V_{bc}  ) c^\dagger_{\bm{k}} c_{\bm{k}} \Big] \nonumber \\
&  + \hbar \omega  a^\dagger a + \sum_{\bm{k}} \left[  g  c^\dagger_{\bm{k}}  h^\dagger_{-\bm{k}}  a 
+  g^* h_{-\bm{k}} c_{\bm{k}} a^\dagger  \right]   \nonumber \\
& - V_{bc} \sum_{\bm{k} \bm{k}'} \Big[  h^\dagger_{\bm{k}} h_{\bm{k}} c^\dagger_{\bm{k}'} c_{\bm{k}'} 
+   c^\dagger_{\bm{k}} h_{-\bm{k}}^\dagger h_{-\bm{k}'} c_{\bm{k}'} \Big] .
\label{holeyham}
\end{align}
Next, we perform a mean-field expansion with respect to the operators $ c^\dagger_{\bm{k}}  h^\dagger_{-\bm{k}}, a,  h^\dagger_{\bm{k}} h_{\bm{k}} $ and $ c^\dagger_{\bm{k}} c_{\bm{k}} $ and keep only linear terms.  The mean-field approximated Hamiltonian is
\begin{align}
H & - \mu n_{\text{ex}} = E_0  + (\hbar \omega -\mu)  a^\dagger a + g D a + g^* D^* a^\dagger  \nonumber \\
& + \sum_{\bm{k}} \left[ -\xi_{\bm{k}}  h^\dagger_{\bm{k}} h_{\bm{k}}  + 
\xi_{\bm{k}}'  c^\dagger_{\bm{k}} c_{\bm{k}} - \Delta  c^\dagger_{\bm{k}}  h^\dagger_{- \bm{k}}
- \Delta^*    h_{- \bm{k}} c_{\bm{k}} \right]
\label{meanfieldham}
\end{align}
where we defined $ \xi_{\bm{k}}  = \epsilon_{\bm{k}} + V_{bc} X $, $ \xi_{\bm{k}}'  = E_c + \epsilon_{\bm{k}}'  + (N +1- X) V_{bc} - \mu $, $ \Delta  = V_{bc} D - g \lambda^* $, and 
$ E_0 = V_{bc} (X^2 + |D|^2) - 2 \text{Re} ( g D \lambda) +  \sum_{\bm{k}} \epsilon_{\bm{k}} $. The mean field values are $ X = \sum_{\bm{k}} \langle h^\dagger_{\bm{k}} h_{\bm{k}} \rangle = 
\sum_{\bm{k}} \langle c^\dagger_{\bm{k}} c_{\bm{k}} \rangle $, $ D = \sum_{\bm{k}} \langle c^\dagger_{\bm{k}} h^\dagger_{-\bm{k}} \rangle $, $ \lambda = \langle a \rangle $.  The number of excited states and holes are equal due to the symmetry of the Hamiltonian.  We have added a chemical potential term which controls the number of excitations in the system.   The photon part of the Hamiltonian can be diagonalized by making a transformation $ B = a - \lambda $, which yields the condition
\begin{align}
\lambda = - \frac{g D}{\hbar \omega - \mu} .
\end{align}

The fermionic part of the Hamiltonian is equivalent to a BCS Hamiltonian, and can be diagonalized in the same way.  The ground state corresponds to a BCS wavefunction
\begin{align}
| \Phi \rangle = \prod_{\bm{k}} \left[ u_{\bm{k}} + v_{\bm{k}} c^\dagger_{\bm{k}} h^\dagger_{-\bm{k}} \right] |E_0 (k_F) \rangle  ,
\end{align}
where $ |u_{\bm{k}}|^2 +  |v_{\bm{k}}|^2 = 1 $ and $ |E_0 (k_F) \rangle = \prod_{| \bm{k}|  < k_F}  b^\dagger_{\bm{k}} |0 \rangle $ is the Fermi sea.  In order that this state is the ground state, application of (\ref{meanfieldham}) on the above state must give an eigenstate.  This gives the constraint
\begin{align}
2  \delta \xi_{\bm{k}}  u_{\bm{k}} v_{\bm{k}}- \Delta ( |v_{\bm{k}}|^2 -  |u_{\bm{k}}|^2 ) = 0 ,
\end{align}
where $ \delta \xi_{\bm{k}} = (\xi_{\bm{k}}'  - \xi_{\bm{k}})/2 $.  Solving the quadratic equation we find the BCS parameters must obey
\begin{align}
v_{\bm{k}}^2 & = \langle c^\dagger_{\bm{k}} c_{\bm{k}} \rangle = \frac{1}{2} \left(
1- \frac{ \delta \xi_{\bm{k}}}{ E_{\bm{k}}} \right) 
 \\
u_{\bm{k}} v_{\bm{k}} & = \langle c^\dagger_{\bm{k}} h^\dagger_{-\bm{k}} \rangle = \frac{\Delta}{2 E_{\bm{k}}} ,
\end{align}
where $ E_{\bm{k}} = \sqrt{ (\delta \xi_{\bm{k}})^2 + \Delta^2} $.  The fully diagonalized Hamiltonian then is given by 
\begin{align}
H & - \mu n_{\text{ex}} = \epsilon_0 + ( \hbar \omega - \mu) B^\dagger B + 
\sum_{\bm{k}}  E_{\bm{k}} ( \gamma_{\bm{k} 0}^\dagger \gamma_{\bm{k} 0} +  \gamma_{\bm{k} 1}^\dagger \gamma_{\bm{k} 1}  ),
\end{align}
where $ \epsilon_0 = \sum_{\bm{k}} [ \epsilon_{\bm{k}} - E_{\bm{k}}  + \delta \xi_{\bm{k}} ] + 
V_{bc}( |D|^2 + |X|^2) + \frac{ |g D|^2}{\hbar \omega - \mu} $ and the Bogoliubov transformed operators are 
$c_{\bm{k}} = u_{\bm{k}}^*  \gamma_{\bm{k} 0}  + v_{\bm{k}}    \gamma_{\bm{k} 1}^\dagger, 
h_{-\bm{k}}^\dagger = -v_{\bm{k}}^*  \gamma_{\bm{k} 0}  + u_{\bm{k}}    \gamma_{\bm{k} 1}^\dagger $.  

The BCS equations are solved in a self-consistent manner, demanding that 
\begin{align}
\delta \xi_{\bm{k}} & = \frac{E_c}{2} + \frac{ \epsilon_{\bm{k}}' - \epsilon_{\bm{k}}}{2} + V_{bc} ( \frac{N+1}{2} - X ) - \frac{\mu}{2} \nonumber \\
\Delta &  = \left( V_{bc} + \frac{ |g|^2}{\hbar \omega - \mu} \right) D .
\label{deltadef}
\end{align}
The mean-field values $ X = \sum_{\bm{k}} v_{\bm{k}}^2 $, $ D = \sum_{\bm{k}} u_{\bm{k}} v_{\bm{k}}  $ depend on $\delta \xi_{\bm{k}} $ and $ \Delta $ so that by self-consistent iteration one obtains the ground state parameters.

\section{The polariton operator}

First let us write the ground state for the Hamiltonian (2).   In the zero excitation sector $  n_{\text{ex}} | \psi \rangle = 0 $, the zero temperature ground state consists of a Fermi sea 
\begin{align}
|E_0 (k_F) \rangle = \prod_{| \bm{k}|  < k_F}  b^\dagger_{\bm{k}} |0 \rangle,
\label{groundstate}
\end{align}
with energy $ E_0(k_F) = \sum_{| \bm{k}|  < k_F} \epsilon_{\bm{k}} $, and $ k_F $ is the Fermi momentum defined as the magnitude of the momentum satisfying $  \epsilon_{\bm{k}} = \mu_{\text{at}} $. 

We examine the special case where the dispersion of the ground and excited states are the same $ \epsilon_{\bm{k}}' = \epsilon_{\bm{k}} $.  We show in this section that in this case the polariton operator
\begin{align}
p^\dagger  = u a^\dagger + v \frac{1}{\sqrt{N}} \sum_{\bm{k}}  c^\dagger_{\bm{k}}  b_{\bm{k}},
\end{align}
creates a single particle excitation of the Hamiltonian (2).  Here $u, v $ are coefficients that satisfy $ |u|^2 + |v|^2 = 1 $.  Applying the polariton operator to the ground state (\ref{groundstate}) produces the state 
\begin{align}
|p\rangle & \equiv p^\dagger |E_0 (k_F) \rangle = u | a \rangle + v |e \rangle  \nonumber \\
| a \rangle & = a^\dagger \prod_{| \bm{k}|  < k_F}  b^\dagger_{\bm{k}} |0 \rangle \nonumber \\
| e \rangle & = \frac{1}{\sqrt{N}} \sum_{| \bm{k}|  < k_F} c^\dagger_{\bm{k}} \prod_{| \bm{k}'|  < k_F;\bm{k}' \ne \bm{k} }  
b^\dagger_{\bm{k}'} |0 \rangle  .
\label{polaritonstate}
\end{align}
Applying the Hamiltonian (2) to the normalized states $ | a \rangle, | e \rangle $ produces a $ 2 \times 2 $ matrix 
\begin{align}
H_p = \left(
\begin{array}{cc}
E_0(k_F) + \hbar \omega & g \sqrt{N} \nonumber \\
g^* \sqrt{N} & E_0(k_F) + E_c 
\end{array}
\right),
\end{align}
which can be diagonalized to give the coefficients 
\begin{align}
u & = \frac{ E_c - \hbar \omega  \pm \sqrt{(E_c - \hbar \omega)^2 + 4|g|^2 N } }{\sqrt{\cal N}} \nonumber \\
v & =  - \frac{2 g \sqrt{N} }{\sqrt{\cal N}} 
\end{align}
where $ \cal N $ is a suitable normalization factor.  The excitation energy is
\begin{align}
E_p =  \frac{E_c + \hbar \omega}{2} \mp \sqrt{\frac{(E_c - \hbar \omega)^2}{4} + |g|^2 N } ,
\end{align}
The two solutions are called the lower and upper polariton respectively, and are eigenstates of the Hamiltonian (2) according to
\begin{align}
H | p \rangle = (E_0(k_F) + E_p) | p \rangle .  
\end{align}

\section{High-excitation limit}

In this section we solve model in the limit that the photon population in the cavity is very large. 
In this limit, the photons can be treated classically, i.e. $ a \rightarrow \lambda = \langle a \rangle $. 
Starting from (2), we assume that the photon population is large enough that the interaction terms may be ignored $ \lambda \gg V_{bc} $.  The effective Hamiltonian for the fermions can be written as
\begin{align}
H_{\lambda} & = \sum_{\bm{k}} \left[ \epsilon_{\bm{k}}  b^\dagger_{\bm{k}} b_{\bm{k}} + (E_c - \hbar \omega   + \epsilon_{\bm{k}}') c^\dagger_{\bm{k}} c_{\bm{k}} \right] \nonumber \\
& + \sum_{\bm{k}} \left[  g  \lambda c^\dagger_{\bm{k}}  b_{\bm{k}}   + g^*  \lambda^*  b_{\bm{k}}^\dagger c_{\bm{k}} \right]   ,
\label{largelambdaham}
\end{align}
where we have made a transformation of the operators to the rotating frame $ c^\dagger_{\bm{k}} \rightarrow e^{-i\omega t } c^\dagger_{\bm{k}} $.   This describes normal mode splitting of the cavity resonance frequencies \cite{agarwal1984vacuum}.  

The diagonalized operators for (\ref{largelambdaham}) can be defined as
\begin{align}
d_{\bm{k}}^\dagger = u_{\bm{k}} b_{\bm{k}}^\dagger + v_{\bm{k}} c_{\bm{k}}^\dagger,
\end{align}
where $ |u_{\bm{k}} |^2 +  |v_{\bm{k}} |^2 = 1 $. These can be solved to give the values 
\begin{align}
u_{\bm{k}} & = \frac{E_c - \hbar \omega + \epsilon_{\bm{k}}' -\epsilon_{\bm{k}}   \pm \sqrt{(E_c - \hbar \omega +\epsilon_{\bm{k}}' -\epsilon_{\bm{k}} )^2 + 4 |g \lambda|^2 } }{\sqrt{\cal N'}} \nonumber \\
v_{\bm{k}} & = - \frac{2 g \lambda}{\sqrt{\cal N'}}
\end{align}
where $ \cal N' $ is a suitable normalization factor.  Reinstating the photonic energy energies are
\begin{align}
E_{\bm{k}}^\lambda = \frac{E_c+ \hbar \omega+ \epsilon_{\bm{k}}' + \epsilon_{\bm{k}}}{2} \pm \sqrt{ \frac{(E_c - \hbar \omega + \epsilon_{\bm{k}}' -\epsilon_{\bm{k}})^2 }{4} +  |g  \lambda|^2} , 
\end{align}
which is the same as (\ref{brightcavitylimit}).  The ground state then is described by a Fermi sea of the $ d_{\bm{k}} $ states
\begin{align}
|E^\lambda (k_F) \rangle = \prod_{| \bm{k}|  < k_F}  \left[ u_{\bm{k}} b_{\bm{k}}^\dagger + v_{\bm{k}} c_{\bm{k}}^\dagger \right] |0 \rangle
\end{align}
with energy  $ E^\lambda (k_F) = \sum_{| \bm{k}|  < k_F} E_{\bm{k}}^\lambda $. This state can be equivalently be written
\begin{align}
|E^\lambda (k_F) \rangle  = \prod_{\bm{k}} \left[ u_{\bm{k}} + v_{\bm{k}} c^\dagger_{\bm{k}} h^\dagger_{-\bm{k}} \right] |E_0 (k_F) \rangle ,
\end{align}
which is exactly the form of the BCS wavefunction.

\section{Spontaneous emission for an exciton}

In this section we derive the spontaneous emission rate for an exciton state
\begin{align}
| \text{ex}, 0 \rangle = \frac{1}{\sqrt{N}} \sum_{\bm{k}} v_{\bm{k}} c^\dagger_{\bm{k}}  h_{-\bm{k}}^\dagger |E_0 (k_F) \rangle \otimes | 0 \rangle ,
\label{initstateexc}
\end{align}
where $ | 0 \rangle $ denotes the vacuum of the electromagnetic field.  Following Weisskopf-Wigner theory \cite{scully1999quantum}, we assume that such an excited state is unstable against decay by emitting a photon in a continuum of electromagnetic fields.  The Hamiltonian reads
\begin{align}
H_{\text{W}}/\hbar  =&  \omega_{\text{ex}} | \text{ex}, 0 \rangle  \langle \text{ex}, 0 |
+ \sum_{\bm{q},s}  \omega_{\bm{q}} a_{\bm{q}s}^\dagger a_{\bm{q}s} \nonumber \\
& - \sum_{\bm{q},s} \sum_{\bm{k}}\left( 
 g_{\bm{q}s} c^\dagger_{\bm{k}} h_{-\bm{k}}^\dagger  a_{\bm{q}s} 
 +  g_{\bm{q}s}^*  h_{-\bm{k}} c_{\bm{k}} a_{\bm{q}s}^\dagger \right)
\label{hamw}
\end{align}
where the atom-field coupling coefficient is
\begin{align}
g_{\bm{q}s} = i  \bm{d}\cdot \bm{\epsilon_{\bm{q}s}}  \sqrt{\frac{\omega_{\bm{q}}}{2 \hbar \epsilon_0 V}}.
\end{align}
Here $ \hbar \omega_{\text{ex}}  $ is the exciton energy, $ \hbar  \omega_{\bm{q}} $ is the energy of the electromagnetic field, $ \epsilon_0 $ is the permittivity, $ \bm{d} $ is the dipole moment of the atom, $  \bm{\epsilon_{\bm{q}s}} $ are the polarization vectors $s\in\{1,2 \} $.    

Given that (\ref{initstateexc}) is the initial state, the Hamiltonian evolves the system according to 
\begin{align}
| \psi(t) \rangle = \chi(t) e^{- i \omega_{\text{ex}} t }   | \text{ex}, 0 \rangle +
\sum_{\bm{q},s} \xi_{\bm{q} s}(t)  e^{-i  \omega_{\bm{q}} t} |E_0 (k_F) , 1_{\bm{q}s} \rangle 
\label{psistatechi}
\end{align}
where
\begin{align}
|E_0 (k_F) ,  1_{\bm{q}s} \rangle  = |E_0 (k_F) \rangle \otimes a_{\bm{q}s}^\dagger  | 0 \rangle .
\end{align}

Substituting (\ref{psistatechi}) and (\ref{hamw}) into the Schrodinger equation
\begin{align}
i \hbar \frac{d | \psi(t) \rangle }{dt} = H_W | \psi(t) \rangle ,
\end{align}
we then have
\begin{align}
& i \hbar e^{-i \omega_{\text{ex}}  t } \frac{d \chi}{dt} | \text{ex}, 0 \rangle  \nonumber \\
& + i \hbar \sum_{\bm{q},s} e^{-i  \omega_{\bm{q}} t} \frac{d \xi_{\bm{q} s} }{dt}  |E_0 (k_F) , 1_{\bm{q}s} \rangle  =  \nonumber \\
& - \chi(t) e^{-i \omega_{\text{ex}}  t }  \left( 
 \frac{1}{\sqrt{N}} \sum_{\bm{k}} v_{\bm{k}} \right) \sum_{\bm{q},s} \hbar g_{\bm{q}s}^* |E_0 (k_F) , 1_{\bm{q}s} \rangle \nonumber \\
& - \sum_{\bm{q},s}  \hbar g_{\bm{q}s}  \xi_{\bm{q} s}(t)  e^{-i  \omega_{\bm{q}} t}  
 \sum_{\bm{k}} c^\dagger_{\bm{k}}  h_{-\bm{k}}^\dagger |E_0 (k_F), 0 \rangle .
\label{schrodingerman}
\end{align}
Multiplying (\ref{schrodingerman}) by $ \langle \text{ex}, 0 | $, we have 
\begin{align}
\frac{d \chi}{dt} = i   \left( \frac{1}{\sqrt{N}}  \sum_{\bm{k}} v_{\bm{k}}^* \right)  \sum_{\bm{q},s}  g_{\bm{q}s}  \xi_{\bm{q} s}(t)  e^{-i  (\omega_{\bm{q}} - \omega_{\text{ex}}) t}  .
\label{dchi}
\end{align}
Multiplying (\ref{schrodingerman}) by $ \langle E_0 (k_F) , 1_{\bm{q}s} | $, we have 
\begin{align}
\frac{d \xi_{\bm{q} s} }{dt} = i \left( 
 \frac{1}{\sqrt{N}} \sum_{\bm{k}} v_{\bm{k}} \right) g_{\bm{q}s}^*    \chi(t) e^{-i (\omega_{\text{ex}} - \omega_{\bm{q}}) t } . 
\label{dxi}
\end{align}
Combining (\ref{dchi}) and (\ref{dxi}), we obtain
\begin{align}
\frac{d \chi}{dt} = -  \left| \frac{1}{\sqrt{N}}  \sum_{\bm{k}} v_{\bm{k}} \right|^2  \sum_{\bm{q},s} | g_{\bm{q}s}  |^2 \int_0^t \chi(t') 
 e^{-i (\omega_{\text{ex}} - \omega_{\bm{q}}) (t'-t) } dt' .  
\end{align}
Under the Markov approximation and assuming that the variation of $ \chi (t) $ is much slower than the rate $ \omega_{\text{ex}} $, we can take 
\begin{align}
\frac{d \chi}{dt} & \approx -  \left| \frac{1}{\sqrt{N}}  \sum_{\bm{k}} v_{\bm{k}} \right|^2  \sum_{\bm{q},s} | g_{\bm{q}s}  |^2 \chi(t)  \int_0^\infty 
 e^{-i (\omega_{\text{ex}} - \omega_{\bm{q}}) (t'-t) } dt'  \nonumber \\
& = -  \left| \frac{1}{\sqrt{N}}  \sum_{\bm{k}} v_{\bm{k}} \right|^2  \pi
| g_{\bm{q}_{\text{ex}}  s} |^2 \chi(t) 
\end{align}
since $ \int_0^\infty  e^{-i (\omega_{\text{ex}} - \omega_{\bm{q}}) t' } dt' = \pi \delta (\omega_{\text{ex}} - \omega_{\bm{q}} ) $, and $ \bm{q}_{\text{ex}} $ is the momentum associated with $  \omega_{\text{ex}} =  \omega_{\bm{q}} $.  

A standard calculation of the coupling yields 
\begin{align}
| g_{\bm{q}_{\text{ex}}  s} |^2 = \frac{ | \bm{d} |^2 \omega_{\text{ex}}^3}{6 \pi^2 \epsilon_0 \hbar c^3}.  
\end{align}
The decay of the amplitude $ \chi(t) $ is then written
\begin{align}
\frac{d \chi}{dt} = - \frac{\Gamma}{2} \chi(t)
\end{align}
where the spontaneous emission rate is
\begin{align}
\Gamma =  \Gamma_0  \left| \frac{1}{\sqrt{N}}  \sum_{\bm{k}} v_{\bm{k}} \right|^2  .
\end{align}
and 
\begin{align}
\Gamma_0 = \frac{ | \bm{d} |^2 \omega_{\text{ex}}^3}{3 \pi \epsilon_0 \hbar c^3} .
\end{align}
To a very good approximation the exciton frequency is the same as the atomic transition energy $ \omega_{\text{ex}} \approx \omega_0 $.  Therefore we can approximate
\begin{align}
\Gamma_0 \approx  \frac{ | \bm{d} |^2 \omega_{0}^3}{3 \pi \epsilon_0 \hbar c^3} ,
\end{align}
which is the single atom spontaneous emission rate.

\section{Spectrum of non-interacting fermions in a cavity}

In this section we derive the spectrum of the two-level fermion under photonic driving and cavity loss.  We are particularly interested in how the spectrum evolves in the high excitation limit, since in the low excitation limit it is well-established that vacuum Rabi splitting is observed.   In this limit, we expect that the physics is dominated by the photon-atom coupling term (i.e. the third term of Eq. (\ref{ham})), and the interactions and the difference in the ground and excited state dispersions do not play a major role.  This is because in the high excitation regime the BCS solution always approaches $ u_{\bm{k}} = v_{\bm{k}} = 1/\sqrt{2} $ independent of the interactions and dispersion.  We first derive the effective Tavis-Cummings model accounting for driving and loss, and then show the master equation from which the cavity spectrum can be found. 

\subsection{Effective Tavis-Cummings model}

We start with (\ref{ham}) neglecting the interaction terms and setting $ \epsilon_{\bm{k}}' = \epsilon_{\bm{k}} $ gives
\begin{align}
H_C & = \sum_{\bm{k}} \left[ \epsilon_{\bm{k}}  b^\dagger_{\bm{k}} b_{\bm{k}} + (E_c + \epsilon_{\bm{k}}) c^\dagger_{\bm{k}} c_{\bm{k}} \right]  + \hbar \omega  a^\dagger a  \nonumber \\
& +\sum_{\bm{k}} \left[    g  c^\dagger_{\bm{k}}  b_{\bm{k}}  a +  g^* b_{\bm{k}}^\dagger c_{\bm{k}} a^\dagger  \right]  .
\end{align}
We now define effective spin operators as
\begin{align}
S^- & =  \sum_{\bm{k}} \sigma^-_{\bm{k}}  =  \sum_{\bm{k}}  b_{\bm{k}}^\dagger c_{\bm{k}}  \nonumber \\
S^+ & =  \sum_{\bm{k}} \sigma^+_{\bm{k}}  =  \sum_{\bm{k}} c^\dagger_{\bm{k}}  b_{\bm{k}}  \nonumber \\
S^z & =  \sum_{\bm{k}} \sigma^z_{\bm{k}}  =  \sum_{\bm{k}} b_{\bm{k}}^\dagger b_{\bm{k}} -  c^\dagger_{\bm{k}}  c_{\bm{k}} .
\end{align}
The Hamiltonian is then written
\begin{align}
H_C & = E_0 + \frac{E_c}{2} S^z + \hbar \omega a^\dagger a +  g S^+ a + g^* S^- a^\dagger 
\label{taviscummings}
\end{align}
where
\begin{align}
E_0 =  \sum_{\bm{k}} \left( \epsilon_{\bm{k}} + \frac{E_c}{2} \right) .
\end{align}
This takes the form of a Tavis-Cummings model. 

Driving the cavity corresponds to adding a photon displacement term
\begin{align}
H_P = \hbar \xi ( a^\dagger e^{-i \gamma t} + a e^{i \gamma t} ) ,
\end{align}
where $ \xi $ is the drive amplitude and $ \gamma $ is the phase of the coherent light.  The master equation for the system is then given by
\begin{align}
\frac{d \rho}{dt} = - \frac{i}{\hbar} [ H_C + H_P, \rho ] + \kappa {\cal D} [a] \rho,
\label{mastertavis}
\end{align}
where the loss of the photons from the cavity is taken into account by the Lindblad superoperator
\begin{align}
{\cal D} [A] \rho = A \rho A^\dagger - \frac{1}{2} A^\dagger  A \rho - \frac{1}{2} \rho  A^\dagger  A .
\end{align}

Following Ref. \cite{bishop2009nonlinear}, we derive the spectrum of the cavity by evaluating the steady-state expectation value $ | \langle a \rangle |^2 $. This is performed by directly evolving the master equation (\ref{mastertavis}) for a sufficiently long time such that $ | \langle a \rangle |^2 $ stabilizes.   The total Hamiltonian $ H_C + H_P $ conserves the atom number
\begin{align}
b^\dagger b + c^\dagger c = N ,
\end{align}
but the photon number is not conserved due to the pump and loss.  We impose a maximum number of photons $ n_{\max} $ in the density matrix and calculate results for sufficiently large $ n_{\max} $ such that the results converge.


\end{document}